\documentclass{article}

\usepackage{arxiv}

\usepackage[utf8]{inputenc} 
\usepackage[T1]{fontenc}    
\usepackage{hyperref}       
\usepackage{url}            
\usepackage{booktabs}       
\usepackage{amsfonts}       
\usepackage{nicefrac}       
\usepackage{microtype}      
\usepackage{lipsum}		
\usepackage{graphicx}
\usepackage{natbib}
\usepackage{doi}
\usepackage{dcolumn}
\usepackage{bm}
\usepackage{siunitx}
\usepackage{authblk}

\title{Suppression of Electromagnetic Crosstalk by Differential Excitation for SAW Generation}

\author[1,2]{Shunsuke Ota}
\author[2]{Yuma Okazaki}
\author[2]{Shuji Nakamura}
\author[2]{Takehiko Oe}
\author[3]{Hermann Sellier}
\author[3]{Christopher B\"auerle}
\author[2]{Nobu-Hisa Kaneko}
\author[1]{Tetsuo Kodera}
\author[2,4]{Shintaro Takada}

\affil[1]{Department of Electrical and Electronic Engineering, Tokyo Institute of Technology, Tokyo 152-8550, Japan}
\affil[2]{National Institute of Advanced Industrial Science and Technology (AIST), National Metrology Institute of Japan (NMIJ), 1-1-1 Umezono, Tsukuba, Ibaraki 305-8563, Japan}
\affil[3]{Univ. Grenoble Alpes, CNRS, Grenoble INP, Institut N\'eel, 38000 Grenoble, France}
\affil[4]{Persent address: Department of Physics, Graduate School of Science, Osaka University, Toyonaka, Osaka 560-0043, Japan}
\affil[ ]{\textit {Corresponding author: ota.s.ab@m.titech.ac.jp}}

\renewcommand{\shorttitle}{Suppression of Electromagnetic Crosstalk by Differential Excitation for SAW Generation}

\begin{document}
\maketitle

\begin{abstract}
Surface acoustic waves (SAWs) hold a vast potential in various fields such as spintronics, quantum acoustics, and electron-quantum optics, but an electromagnetic wave emanating from SAW generation circuits has often been a major hurdle. Here, we investigate a differential excitation method of interdigital transducers (IDTs) to generate SAWs while reducing the electromagnetic wave. The results show that electromagnetic waves are suppressed by more than \SI{90}{\%} in all directions. This suppression overcomes the operating limits and improves the scalability of SAW systems. Our results promise to facilitate the development of SAW-based applications in a wide range of research fields.
\end{abstract}


Surface acoustic waves (SAWs) are a versatile phononic technology widely used in both industrial applications and basic research \cite{Delsing2019}. In industry, they are crucial for frequency filtering in wireless devices \cite{IEEE2017} and are used for the sensing and mixing of trace amounts of liquids \cite{Go2017}. In the realm of basic research, SAWs are leading to increasingly impressive results in spintronics, such as spin currents \cite{Kobayashi2017} and skyrmion generation \cite{Yokouchi2020}, and in quantum acoustics, such as coupling to superconducting qubits \cite{Gustafsson2014,Manenti2017,Noguchi2017,Moores2018,Bolgar2018} and developing a beam splitter \cite{Qiao2023}. Furthermore, in the field of electron-quantum optics, various studies have been conducted \cite{Hermelin2011,McNeil2011,Stotz2005, Bertrand2016,Takada2019,Jadot2021,Ito2021,Edlbauer2021,Wang2022,Wang2023}. In combination with quantum dots on semiconductors, the transport of single electrons across micrometer distances has been demonstrated \cite{Hermelin2011,McNeil2011,Takada2019}, as well as the coherent transport of electron spins \cite{Bertrand2016,Jadot2021}. The use of SAWs in quantum information processing has been proposed for some time \cite{Barnes2000,Foden2000,Roberta2005}, and recently the development of electron flying qubits \cite{Bäuerle2018,Edlbauer2022} has been studied.

Meanwhile, such developments of SAWs often face the persistent problem of electromagnetic wave crosstalk. High frequency components in experimental circuits to generate SAWs emit electromagnetic waves while generating SAWs. This electromagnetic wave is picked up by the metal gates of the target structure, which are used to conduct the intended experiments (e.g., the metal gates that define the quantum dots), and generates fluctuation of the electric potential in the target. This undesired fluctuation not only hides the desired effect by the SAW, but also interferes with the SAW to produce a negative effect \cite{Kataoka2006,Ota2023}. Indeed, this crosstalk is pronounced and should be one of the reasons why quantum current sources using SAWs were eventually largely abandoned, despite great efforts over the years \cite{Shilton1996,Talyanskii1997,Cunningham1999,Cunningham2000,Ebbecke2002,Utko2003,Ford2017}.

In the field of electron-quantum optics, the crosstalk problem has been avoided by shifting the arrival timing of the electromagnetic wave and the SAW at the target structure where electrons are transported.
This could be done by making the length of the SAW generation signal sufficiently shorter than the propagation time of the SAW from the generated position to the target structure.
Since the electromagnetic wave propagates with the speed of light (\SI{3e8}{m/s}), which is 5 orders of magnitude faster than the speed of SAWs ($\sim$\SI{3000}{m/s}), it disappears immediately after we stop applying the SAW generation signal and before the SAW arrives at the target structure.
Although this method allows for avoiding critical influence of the crosstalk, it severely limits the timing of SAW generation and continuous generation of SAW is prohibited.
This inflexibility is a clear disadvantage for various applications.
For example, in electron-quantum optics, it makes it difficult to scale up the system, where many single electrons are transferred by SAWs at different timings.
In this study, we investigate the way not to avoid the crosstalk but to suppress it by devising the SAW generation method.
With the developed method we have succeeded in radically reducing the electromagnetic crosstalk.

\begin{figure}[t]
 \begin{center}
  \includegraphics[width=0.5\textwidth]{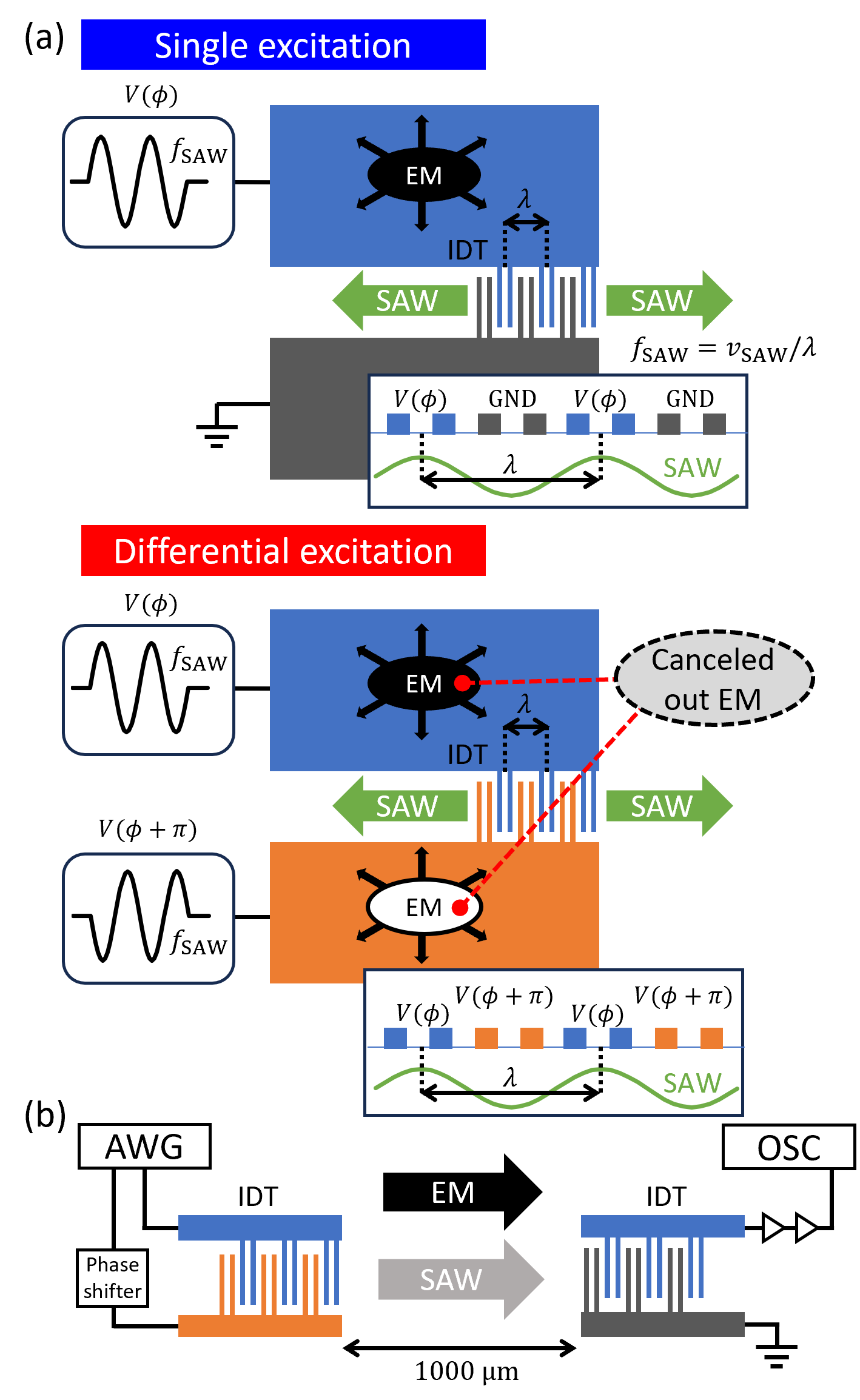}
  \caption{(a) Schematic diagram of the excitation method. Top: single excitation, bottom: differential excitation. The SAW generation signal $V(\phi)$ is input into the pads connecting to the comb-like electrodes of the IDT, and the SAWs and electromagnetic waves (EM) are generated. In the case of the single excitation, the electromagnetic wave originated from $V(\phi)$ is emitted from the high-frequency circuit such as metal pads. In the case of the differential excitation, the two types of electromagnetic waves originated from $V(\phi)$ and $V(\phi+\pi)$ destructively interfere with each other and cancel out. (b) Schematic diagram of the experimental setup and the device for the differential excitation.}
  \label{figure1}
 \end{center}
\end{figure}

SAWs are generated by using comb-shaped electrodes called an interdigital transducer (IDT) on a piezoelectric substrate.
Top part of Fig.\,\ref{figure1}a shows a typical scheme to generate SAWs, where one pad of the IDT is grounded while the other is excited by an ac voltage.
In this study, this scheme is called a single excitation.
Here, IDTs with a double-finger pattern are employed.
This type of IDT is often employed because it suppresses reflections of SAWs inside the IDT and improves conduction efficiency.
When we apply an ac voltage whose frequency corresponds to the resonance frequency, $f_{\rm SAW}$, determined by the period of the IDT fingers, $\lambda$, and the SAW velocity of the substrate, $V_{\rm SAW}$, as $f_{\rm SAW} = v_{\rm SAW}/ \lambda$ on the IDT, SAWs generated from each finger constructively interfere and strong SAWs propagate along the substrate towards the both directions.
In this process, electromagnetic waves with the same frequency, $f_{\rm SAW}$, are emitted from the metal pad where the excitation voltage is applied.
In this study, we perform a differential excitation of the IDT as shown in the bottom part of Fig.\,\ref{figure1}a to suppress the radiation of the electromagnetic waves.
In the differential excitation, the resonant ac voltage is applied to the both metal pads with a same amplitude but a phase shift of $\pi$ between one to the other.
Since the shape of the applied ac field is same for the differential excitation and the single excitation as depicted in Fig.\,\ref{figure1}a, SAWs are generated as in the case of the single excitation.
Different from the single excitation, electromagnetic waves are emitted from the both pads.
These electromagnetic waves have a phase difference of $\pi$ and the amplitude is expected to be equal if the shapes of the metal pads are symmetrical.
Since the wavelength of the electromagnetic waves, which is about \SI{100}{mm} for $f_{\rm SAW} =$ \SI{3}{GHz}, are about 100 times longer than the scale of the IDT, which is typically less than \SI{1}{mm} for the IDT with a resonant frequency in the GHz range, the two electromagnetic waves are expected to destructively interfere. As a result, the electromagnetic wave radiated from the IDT should be strongly suppressed.

\begin{figure}[t]
 \begin{center}
  \includegraphics[width=0.7\textwidth]{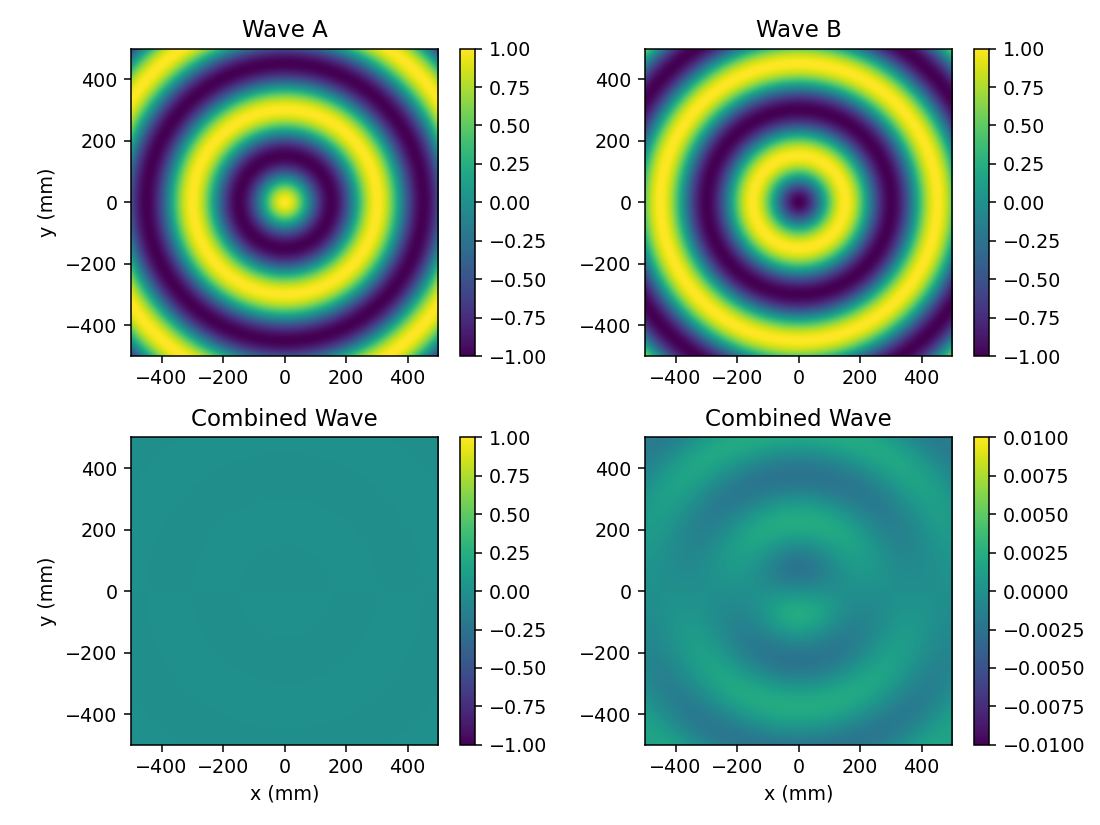}
  \caption{Two-dimensional simulation results of the cancellation effect due to the interference between two circular waves with opposite phases: Wave A and Wave B. The wave centers correspond to metal pads \SI{110}{\micro m} apart on our device, where a resonant ac voltage is applied to excite SAWs. Top left: Wave A - a frequency \SI{1}{GHz}, a phase 0, a velocity \SI{3e8}{m/s}, from point A (\SI{0}{\micro m},\SI{55}{\micro m}). Top right: Wave B - a frequency \SI{1}{GHz}, a phase $\pi$, a velocity \SI{3e8}{m/s}, from point B (\SI{0}{\micro m},\SI{-55}{\micro m}). Lower left: Combined Wave - interference pattern resulting from the combination of Wave A and Wave B, showing the cancelation effect. Lower right: Combined Wave (Detailed View) - Detailed interference pattern with a narrowed amplitude range.}
  \label{figure2}
 \end{center}
\end{figure}

A simple simulation is performed to estimate the expected cancellation effect of electromagnetic waves due to the differential excitation. In this simulation, waves of the same frequency emitted from two different center points are modeled as electromagnetic waves and their interference with each other is calculated. Each center point is assumed to be the center of the metal pads to which the resonant ac voltage is applied. In our sample design, spacing of the two metal pads are \SI{110}{\micro m}. The waves emitted from both center points have a frequency of \SI{1}{GHz}, an equal amplitude of 1, and a velocity of \SI{3e8}{m/s}. The phase of the wave from center point A is set to 0 and the one from center point B is set to $\pi$. This produces waves in perfectly opposite phases. These waves are calculated on a \SI{1}{m} square two-dimensional plane so that the interference pattern could be easily checked. The circular waves are calculated taking into account the phase difference that depends on the distance from the center, and the amplitude is set to a constant value independent of distance. Fig.\,\ref{figure2} shows the respective electromagnetic waves (Wave A and Wave B) in the top panels and the interference pattern resulting from the synthesis of both waves (Combined Wave) in the lower panels. In the lower left panel, the amplitude range of Combined Wave is set from -1 to 1, and the cancellation effect of the combined electromagnetic wave can be clearly observed. In the lower right panel, the amplitude range is set from -0.01 to 0.01 in order to confirm the detailed interference pattern. From this result, the cancellation effect of the electromagnetic waves is estimated to be over \SI{99}{\%}. The reason why a considerable suppression is estimated after taking into account the phase shift due to the difference in emission position of the two electromagnetic waves is that the difference in emission position (\SI{110}{\micro m}) is small enough for the wavelength of the \SI{1}{GHz} electromagnetic wave (\SI{300}{mm}).
Similar results are also obtained at other frequencies commonly used for SAW generation signals (\SI{3}{GHz} - \SI{4}{GHz}).

Next, we perform measurement of the device with a semi-automatic probe station at room temperature. The device is fabricated on a GaAs substrate and consists of IDTs and metal pads connected to the IDTs. The electrodes of the IDTs are fabricated using a standard electron-beam lithography with successive thin-film evaporation (metalization Ti \SI{3}{nm}, Al \SI{27}{nm}). The metal pads are fabricated using a standard photolithography with successive thin-film evaporation (metalization Ti \SI{20}{nm}, Au \SI{100}{nm}). Two IDTs of the same design are placed facing each other in the direction of SAW propagation, at a distance of \SI{1000}{\micro m}, as shown in Fig.\,\ref{figure1}b. One IDT acts as a SAW generator and the other as a SAW detector.
The both IDTs are designed to generate and detect SAWs at a resonant frequency of \SI{1}{GHz} ($\lambda = $\SI{2.86}{\micro m}) and have 40 pairs of electrodes
, an aperture of \SI{30}{\micro m}. The SAW propagation direction is set to [110] direction. Contact metal pads (\SI{330}{\micro m} length and \SI{60}{\micro m} width) are connected to the upper- and the lower-electrode sets of each IDT. High-frequency probes make contact with these pads, thus establishing a connection to the IDTs. These pads also act as a detector of electromagnetic waves. The resonant ac voltage to excite SAWs is provided by an arbitrary waveform generator (AWG, Keysight M8195A) and fed through different coaxial cables to the upper and lower electrodes of the IDT, respectively, in the case of differential excitation.
A phase shifter (WAKA 02X0442-00) is connected to one of the input lines for fine phase difference adjustment before the signal is input to the IDT. The phase shifter is adjusted and fixed so that there is minimal phase displacement at the probe when the same ac voltage is output from two channels of AWG. On the detector side, the set of lower electrodes is grounded (embedded in the surrounding ground pad). The set of upper electrodes is connected to a high-speed sampling oscilloscope (Keysight N1094B DCA-M) to observe the generated SAW and the electromagnetic wave. The detected signal is amplified by a series of broadband amplifiers (SHF S126A, ZHL-4W-422+).

\begin{figure}[t]
 \begin{center}
  \includegraphics[width=0.7\textwidth]{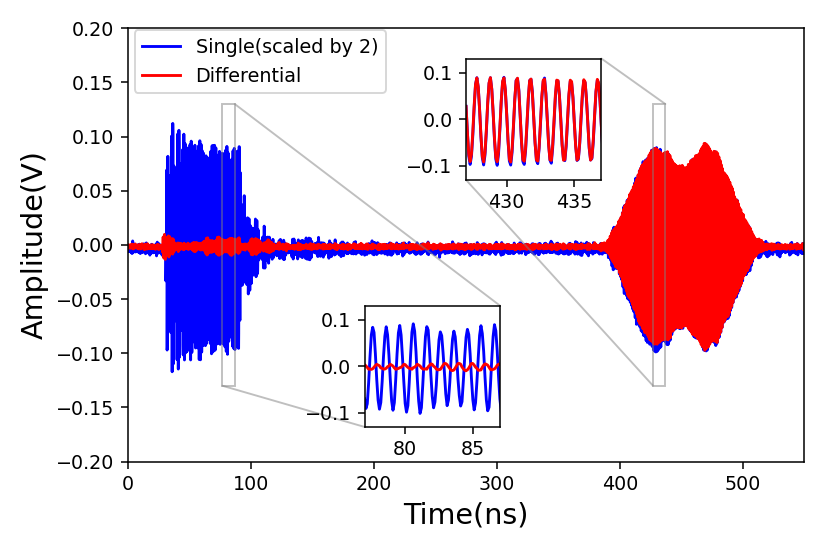}
  \caption{Time-resolved measurements on the IDT with the single excitation and the differential excitation. Trace of the detector response for the resonant ac voltage(\SI{1}{GHz}, a time span \SI{60}{ns}, a peak-to-peak amplitude \SI{350}{mV}). The blue line represents the SAW and the electromagnetic wave signals obtained from the single excitation, with the SAW on the right and the previously detected electromagnetic wave on the left. The red line represents the obtained signal from the differential excitation, where the SAW amplitude is similar to the single excitation while the electromagnetic component is significantly suppressed}.
  \label{figure3}
 \end{center}
\end{figure}
SAWs and electromagnetic waves are generated and observed with the single and differential excitation.
To generate SAWs, we apply a sinusoidal signal at the resonance frequency, \SI{1}{GHz}, with a time span of \SI{60}{ns} and a peak-to-peak amplitude of \SI{350}{mV} in a repetition period of \SI{1520}{ns}.
Here the time span of the signal, \SI{60}{ns}, is chosen to be longer than \SI{40}{ns} which is the minimum length to fully excite the IDT and to be shorter than the propagation time of SAWs between the IDTs, $\sim$ \SI{350}{ns} which is calculated from the distance between the IDTs (\SI{1000}{\micro m}) and the SAW speed in GaAs ($\sim$ \SI{2860}{m/s}).
Since the electromagnetic waves emitted from the IDT propagate with the speed of light ($\sim$ \SI{3e8}{m/s}), they reach the other detector IDT almost instantaneously in less than \SI{10}{ps}.
Therefore, when we perform a real-time detection of the signal at the detector IDT, we first observe the electromagnetic waves and later observe the SAW signal with a well-defined separation between within the repetition period of \SI{1520}{ns}.
Here we apply the signal with the same amplitude for both the single and the differential excitation and hence the detected SAW signal is expected to be double for the differential excitation compared to the single. To compare the ratio between the amplitude of the SAW signal and the electromagnetic waves for both excitation methods we double the detected signal for the single excitation.

Fig.\,\ref{figure3} shows the signal measured with the sampling oscilloscope. The blue line is the data obtained when we perform the single excitation.
The signal observed in the earlier time comes from electromagnetic waves and the one observed in approximately \SI{350}{ns} later comes from SAWs.
For the single excitation, the signal coming from electromagnetic waves is as large as the one coming from SAWs.
The red line shows the data obtained when we perform the differential excitation.
The signal coming from SAWs is as large as the one from the single excitation.
On the other hand, the signal coming from electromagnetic waves is strongly suppressed as expected from our simple simulation.
Comparing the average values of the signal amplitude coming from electromagnetic waves from the single and the differential excitation, it is calculated that \SI{92.4}{\%} of the signal is suppressed for the case of the differential excitation.
In our simple simulation, the electromagnetic-wave suppression of about \SI{99}{\%} is expected.
The residual amount of the electromagnetic waves can be attributed to minute discrepancies between the two emitted electromagnetic waves due to the difference of response functions of the metal pads.
These discrepancies manifested as time-dependence components that exhibited different phases and intensities and could not be addressed by simple adjustments of the phase or the amplitude of the SAW generation signal.
Such fluctuations are considered to be intricately related to high-frequency circuit components like a shape or a thickness of the metal pads.
It suggests that optimization of these components is essential for even better suppression of the electromagnetic waves.

\begin{figure*}[htb]
 \centering
 \includegraphics[width=0.8\textwidth]{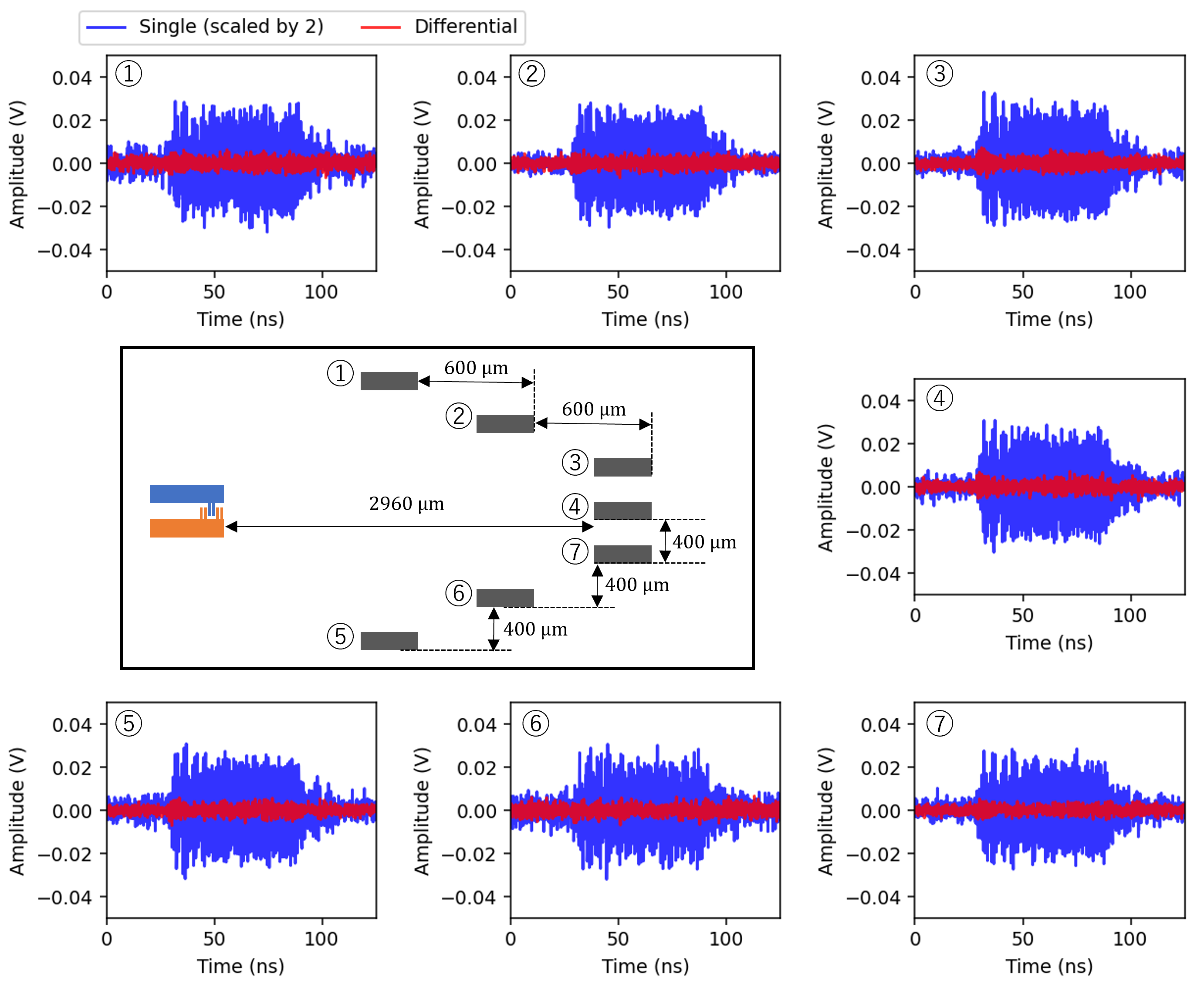}
 \caption{Spatial dependence of electromagnetic-wave suppression. A schematic of the sample is shown in the middle. Surrounding plots are electromagnetic waves detected on each metal pad with the single and the differential excitation. The numbers of the metal pads in the schematic correspond to the numbers of the plots. The suppression effect is almost same for all positions. The change in the amplitude of the electromagnetic waves from the results in Fig.\,\ref{figure3} can be attributed to the different design of the metal pads for detection.}
 \label{figure4}
\end{figure*}
Finally, we experimentally confirm the spatial dependence of the electromagnetic-wave suppression.
For that purpose, we prepared the other device illustrated in the schematic in the middle of Fig.\,\ref{figure4}.
On the left side, an IDT and metal pads having the same design as the device measured in Fig.\,\ref{figure3} are placed.
On the right side, multiple metal pads (\SI{220}{\micro m} in length and \SI{60}{\micro m} in width) are arranged to pick up and detect electromagnetic waves.
They are symmetrically arranged about the horizontal line passing through the middle of the IDT.
The signal used to generate SAWs is the same as in Fig.\,\ref{figure3}.
Each plot in Fig.\,\ref{figure4} shows the signal coming from electromagnetic waves detected by the individual metal pads for the case of the single and the differential excitation.
From these results, it is confirmed that the electromagnetic-wave suppression occurs similarly at all positions of the metal pads.

In conclusion, our study demonstrates the effectiveness of the differential excitation in mitigating electromagnetic crosstalk in SAW-based devices. The experimental results have shown that this approach is capable of nullifying over \SI{90}{\%} of the undesirable electromagnetic components, achieving a substantial reduction.
This suppression of the electromagnetic crosstalk will certainly contribute to the improvement of the accuracy to control single flying electrons using SAW potentials \cite{Kataoka2006}.
Furthermore, the spatial dependence of the electromagnetic wave cancellation effect has also been investigated. Consistent reduction in electromagnetic waves has been evident at all measured positions. This consistency is important for the scalability of SAW-based devices.
In addition, although our experiments have been conducted at room temperature, we anticipate that the principles of the electromagnetic-wave suppression with the differential excitation are equally applicable in low-temperature environments.
The electromagnetic-wave suppression has been investigated with a double-finger IDT in this study. However, it will be equally applicable to various types of IDTs \cite{Morgan2007,Lima2003,Schülein2015,Ekström2017,Dumur2019}.
Our research focuses on fundamentally suppressing the electromagnetic crosstalk that has long hindered advancements of SAW-based single-electron control devices.
The differential excitation of IDTs provides a new perspective for addressing the problem of electromagnetic crosstalk in SAW-based systems and opens up the possibility of application of SAWs in various fields of research such as spintronics and quantum acoustics. For example, in electron-quantum optics, the methodologies and insights obtained in this study are expected to significantly improve the scalability of SAW-based single-electron manipulation devices and make important contributions to the evolution of sophisticated SAW quantum information processing devices.

Acknowledgments \\
S.O. acknowledge financial support from JST SPRING, Grant Number JPMJSP2106.\\
T.K. and S.T. acknowledge financial support from JSPS KAKENHI Grant Number 20H02559. \\
N.-H.K. acknowledges financial support from JSPS KAKENHI Grant Number JP18H05258. \\
C.B. acknowledges financial support from
the French Agence Nationale de la Recherche (ANR), project QUABS ANR-21-CE47-0013-01 as well as funding from the European Union’s H2020 research and innovation program under grant agreement No 862683 ”UltraFastNano”.\\

\bibliographystyle{unsrtnat}
\bibliography{template}  

\begin{thebibliography}{40}
\providecommand{\natexlab}[1]{#1}
\providecommand{\url}[1]{\texttt{#1}}
\expandafter\ifx\csname urlstyle\endcsname\relax
  \providecommand{\doi}[1]{doi: #1}\else
  \providecommand{\doi}{doi: \begingroup \urlstyle{rm}\Url}\fi

\bibitem[Delsing et~al.(2019)Delsing, Cleland, Schuetz, Knörzer, Giedke, Cirac, Srinivasan, Wu, Balram, Bäuerle, Meunier, Ford, Santos, Cerda-Méndez, Wang, Krenner, Nysten, Weiß, Nash, Thevenard, Gourdon, Rovillain, Marangolo, Duquesne, Fischerauer, Ruile, Reiner, Paschke, Denysenko, Volkmer, Wixforth, Bruus, Wiklund, Reboud, Cooper, Fu, Brugger, Rehfeldt, and Westerhausen]{Delsing2019}
P.~Delsing, A.~N. Cleland, M.~J.~A. Schuetz, J.~Knörzer, G.~Giedke, J.~I. Cirac, K.~Srinivasan, M.~Wu, K.~C. Balram, C.~Bäuerle, T.~Meunier, C.~J.~B. Ford, P.~V. Santos, E.~Cerda-Méndez, H.~Wang, H.~J. Krenner, E.~D.~S. Nysten, M.~Weiß, G.~R. Nash, L.~Thevenard, C.~Gourdon, P.~Rovillain, M.~Marangolo, J.-Y. Duquesne, G.~Fischerauer, W.~Ruile, A.~Reiner, B.~Paschke, D.~Denysenko, D.~Volkmer, A.~Wixforth, H.~Bruus, M.~Wiklund, J.~Reboud, J.~M. Cooper, Y.~Fu, M.~S. Brugger, F.~Rehfeldt, and C.~Westerhausen.
\newblock The 2019 surface acoustic waves roadmap.
\newblock \emph{Journal of Physics D: Applied Physics}, 52:\penalty0 353001, 2019.
\newblock \doi{10.1088/1361-6463/ab1b04}.

\bibitem[IEE()]{IEEE2017}
{IEEE Future Networks Technology Roadmap Working Group 2017 IEEE 5G and beyond technology roadmap white paper}.

\bibitem[Go et~al.(2017)Go, Atashbar, Ramshani, and Chang]{Go2017}
David~B. Go, Massood~Z. Atashbar, Zeinab Ramshani, and Hsueh-Chia Chang.
\newblock {Surface acoustic wave devices for chemical sensing and microfluidics: a review and perspective}.
\newblock \emph{Analytical Methods}, 9\penalty0 (28):\penalty0 4112--4134, 2017.
\newblock ISSN 1759-9660.
\newblock \doi{10.1039/c7ay00690j}.

\bibitem[Kobayashi et~al.(2017)Kobayashi, Yoshikawa, Matsuo, Iguchi, Maekawa, Saitoh, and Nozaki]{Kobayashi2017}
D.~Kobayashi, T.~Yoshikawa, M.~Matsuo, R.~Iguchi, S.~Maekawa, E.~Saitoh, and Y.~Nozaki.
\newblock {Spin Current Generation Using a Surface Acoustic Wave Generated via Spin-Rotation Coupling}.
\newblock \emph{Physical Review Letters}, 119\penalty0 (7):\penalty0 077202, 2017.
\newblock ISSN 0031-9007.
\newblock \doi{10.1103/physrevlett.119.077202}.

\bibitem[Yokouchi et~al.(2020)Yokouchi, Sugimoto, Rana, Seki, Ogawa, Kasai, and Otani]{Yokouchi2020}
Tomoyuki Yokouchi, Satoshi Sugimoto, Bivas Rana, Shinichiro Seki, Naoki Ogawa, Shinya Kasai, and Yoshichika Otani.
\newblock {Creation of magnetic skyrmions by surface acoustic waves}.
\newblock \emph{Nature Nanotechnology}, 15\penalty0 (5):\penalty0 361--366, 2020.
\newblock ISSN 1748-3387.
\newblock \doi{10.1038/s41565-020-0661-1}.

\bibitem[Gustafsson et~al.(2014)Gustafsson, Aref, Kockum, Ekström, Johansson, and Delsing]{Gustafsson2014}
Martin~V. Gustafsson, Thomas Aref, Anton~Frisk Kockum, Maria~K. Ekström, Göran Johansson, and Per Delsing.
\newblock {Propagating phonons coupled to an artificial atom}.
\newblock \emph{Science}, 346\penalty0 (6206):\penalty0 207--211, 2014.
\newblock ISSN 0036-8075.
\newblock \doi{10.1126/science.1257219}.

\bibitem[Manenti et~al.(2017)Manenti, Kockum, Patterson, Behrle, Rahamim, Tancredi, Nori, and Leek]{Manenti2017}
Riccardo Manenti, Anton~F. Kockum, Andrew Patterson, Tanja Behrle, Joseph Rahamim, Giovanna Tancredi, Franco Nori, and Peter~J. Leek.
\newblock {Circuit quantum acoustodynamics with surface acoustic waves}.
\newblock \emph{Nature Communications}, 8\penalty0 (1):\penalty0 975, 2017.
\newblock \doi{10.1038/s41467-017-01063-9}.

\bibitem[Noguchi et~al.(2017)Noguchi, Yamazaki, Tabuchi, and Nakamura]{Noguchi2017}
Atsushi Noguchi, Rekishu Yamazaki, Yutaka Tabuchi, and Yasunobu Nakamura.
\newblock {Qubit-Assisted Transduction for a Detection of Surface Acoustic Waves near the Quantum Limit}.
\newblock \emph{Physical Review Letters}, 119\penalty0 (18):\penalty0 180505, 2017.
\newblock ISSN 0031-9007.
\newblock \doi{10.1103/physrevlett.119.180505}.

\bibitem[Moores et~al.(2018)Moores, Sletten, Viennot, and Lehnert]{Moores2018}
Bradley~A. Moores, Lucas~R. Sletten, Jeremie~J. Viennot, and K.~W. Lehnert.
\newblock {Cavity Quantum Acoustic Device in the Multimode Strong Coupling Regime}.
\newblock \emph{Physical Review Letters}, 120\penalty0 (22):\penalty0 227701, 2018.
\newblock ISSN 0031-9007.
\newblock \doi{10.1103/physrevlett.120.227701}.

\bibitem[Bolgar et~al.(2018)Bolgar, Zotova, Kirichenko, Besedin, Semenov, Shaikhaidarov, and Astafiev]{Bolgar2018}
Aleksey~N. Bolgar, Julia~I. Zotova, Daniil~D. Kirichenko, Ilia~S. Besedin, Aleksander~V. Semenov, Rais~S. Shaikhaidarov, and Oleg~V. Astafiev.
\newblock {Quantum Regime of a Two-Dimensional Phonon Cavity}.
\newblock \emph{Physical Review Letters}, 120\penalty0 (22):\penalty0 223603, 2018.
\newblock ISSN 0031-9007.
\newblock \doi{10.1103/physrevlett.120.223603}.

\bibitem[Qiao et~al.(2023)Qiao, Dumur, Andersson, Yan, Chou, Grebel, Conner, Joshi, Miller, Povey, Wu, and Cleland]{Qiao2023}
H.~Qiao, É. Dumur, G.~Andersson, H.~Yan, M.-H. Chou, J.~Grebel, C.~R. Conner, Y.~J. Joshi, J.~M. Miller, R.~G. Povey, X.~Wu, and A.~N. Cleland.
\newblock {Splitting phonons: Building a platform for linear mechanical quantum computing}.
\newblock \emph{Science}, 380\penalty0 (6649):\penalty0 1030--1033, 2023.
\newblock ISSN 0036-8075.
\newblock \doi{10.1126/science.adg8715}.

\bibitem[Hermelin et~al.(2011)Hermelin, Takada, Yamamoto, Tarucha, Wieck, Saminadayar, Bäuerle, and Meunier]{Hermelin2011}
S.~Hermelin, S.~Takada, M.~Yamamoto, S.~Tarucha, A.~D. Wieck, L.~Saminadayar, C.~Bäuerle, and T.~Meunier.
\newblock Electrons surfing on a sound wave as a platform for quantum optics with flying electrons.
\newblock \emph{Nature}, 477:\penalty0 435, 2011.
\newblock \doi{10.1038/nature10416}.

\bibitem[McNeil et~al.(2011)McNeil, Kataoka, Ford, Barnes, Anderson, Jones, Farrer, and Ritchie]{McNeil2011}
R.~P.~G. McNeil, M.~Kataoka, C.~J.~B. Ford, C.~H.~W. Barnes, D.~Anderson, G.~A.~C. Jones, I.~Farrer, and D.~A. Ritchie.
\newblock On-demand single-electron transfer between distant quantum dots.
\newblock \emph{Nature}, 477:\penalty0 439, 2011.
\newblock \doi{10.1038/nature10444}.

\bibitem[Stotz et~al.(2005)Stotz, Hey, Santos, and Ploog]{Stotz2005}
J.~A.~H. Stotz, R.~Hey, P.~V. Santos, and K.~H. Ploog.
\newblock Coherent spin transport through dynamic quantum dots.
\newblock \emph{Nature Materials}, 4:\penalty0 585, 2005.
\newblock \doi{10.1038/nmat1430}.

\bibitem[Bertrand et~al.(2016)Bertrand, Hermelin, Takada, Yamamoto, Tarucha, Ludwig, Wieck, Bäuerle, and Meunier]{Bertrand2016}
B.~Bertrand, S.~Hermelin, S.~Takada, M.~Yamamoto, S.~Tarucha, A.~Ludwig, A.~D. Wieck, C.~Bäuerle, and T.~Meunier.
\newblock Fast spin information transfer between distant quantum dots using individual electrons.
\newblock \emph{Nature Nanotechnology}, 11:\penalty0 672, 2016.
\newblock \doi{10.1038/nnano.2016.82}.

\bibitem[Takada et~al.(2019)Takada, Edlbauer, Lepage, Wang, Mortemousque, Georgiou, Barnes, Ford, Yuan, Santos, Waintal, Ludwig, Wieck, Urdampilleta, Meunier, and Bäuerle]{Takada2019}
S.~Takada, H.~Edlbauer, H.~V. Lepage, J.~Wang, P.-A. Mortemousque, G.~Georgiou, C.~H.~W. Barnes, C.~J.~B. Ford, M.~Yuan, P.~V. Santos, X.~Waintal, A.~Ludwig, A.~D. Wieck, M.~Urdampilleta, T.~Meunier, and C.~Bäuerle.
\newblock Sound-driven single-electron transfer in a circuit of coupled quantum rails.
\newblock \emph{Nature Communications}, 10:\penalty0 4557, 2019.
\newblock \doi{10.1038/s41467-019-12514-w}.

\bibitem[Jadot et~al.(2021)Jadot, Mortemousque, Chanrion, Thiney, Ludwig, Wieck, Urdampilleta, Bäuerle, and Meunier]{Jadot2021}
B.~Jadot, P.-A. Mortemousque, E.~Chanrion, V.~Thiney, A.~Ludwig, A.~D. Wieck, M.~Urdampilleta, C.~Bäuerle, and T.~Meunier.
\newblock Distant spin entanglement via fast and coherent electron shuttling.
\newblock \emph{Nature Nanotechnology}, 16:\penalty0 570, 2021.
\newblock \doi{10.1038/s41565-021-00846-y}.

\bibitem[Ito et~al.(2021)Ito, Takada, Ludwig, Wieck, Tarucha, and Yamamoto]{Ito2021}
R.~Ito, S.~Takada, A.~Ludwig, A.~D. Wieck, S.~Tarucha, and M.~Yamamoto.
\newblock Coherent beam splitting of flying electrons driven by a surface acoustic wave.
\newblock \emph{Physical Review Letters}, 126:\penalty0 070501, 2021.
\newblock \doi{10.1103/physrevlett.126.070501}.

\bibitem[Edlbauer et~al.(2021)Edlbauer, Wang, Ota, Richard, Jadot, Mortemousque, Okazaki, Nakamura, Kodera, Kaneko, Ludwig, Wieck, Urdampilleta, Meunier, Bäuerle, and Takada]{Edlbauer2021}
Hermann Edlbauer, Junliang Wang, Shunsuke Ota, Aymeric Richard, Baptiste Jadot, Pierre-André Mortemousque, Yuma Okazaki, Shuji Nakamura, Tetsuo Kodera, Nobu-Hisa Kaneko, Arne Ludwig, Andreas~D. Wieck, Matias Urdampilleta, Tristan Meunier, Christopher Bäuerle, and Shintaro Takada.
\newblock {In-flight distribution of an electron within a surface acoustic wave}.
\newblock \emph{Applied Physics Letters}, 119\penalty0 (11):\penalty0 114004, 2021.
\newblock ISSN 0003-6951.
\newblock \doi{10.1063/5.0062491}.

\bibitem[Wang et~al.(2022)Wang, Ota, Edlbauer, Jadot, Mortemousque, Richard, Okazaki, Nakamura, Ludwig, Wieck, Urdampilleta, Meunier, Kodera, Kaneko, Takada, and Bäuerle]{Wang2022}
Junliang Wang, Shunsuke Ota, Hermann Edlbauer, Baptiste Jadot, Pierre-André Mortemousque, Aymeric Richard, Yuma Okazaki, Shuji Nakamura, Arne Ludwig, Andreas~D. Wieck, Matias Urdampilleta, Tristan Meunier, Tetsuo Kodera, Nobu-Hisa Kaneko, Shintaro Takada, and Christopher Bäuerle.
\newblock {Generation of a Single-Cycle Acoustic Pulse: A Scalable Solution for Transport in Single-Electron Circuits}.
\newblock \emph{Physical Review X}, 12\penalty0 (3):\penalty0 031035, 2022.
\newblock \doi{10.1103/physrevx.12.031035}.

\bibitem[Wang et~al.(2023)Wang, Edlbauer, Richard, Ota, Park, Shim, Ludwig, Wieck, Sim, Urdampilleta, Meunier, Kodera, Kaneko, Sellier, Waintal, Takada, and Bäuerle]{Wang2023}
J.~Wang, H.~Edlbauer, A.~Richard, S.~Ota, W.~Park, J.~Shim, A.~Ludwig, A.~D. Wieck, H.-S. Sim, M.~Urdampilleta, T.~Meunier, T.~Kodera, N.-H. Kaneko, H.~Sellier, X.~Waintal, S.~Takada, and C.~Bäuerle.
\newblock Coulomb-mediated antibunching of an electron pair surfing on sound.
\newblock \emph{Nature Nanotechnology}, 18:\penalty0 721, 2023.
\newblock \doi{10.1038/s41565-023-01368-5}.

\bibitem[Barnes et~al.(2000)Barnes, Shilton, and Robinson]{Barnes2000}
C.~H.~W. Barnes, J.~M. Shilton, and A.~M. Robinson.
\newblock {Quantum computation using electrons trapped by surface acoustic waves}.
\newblock \emph{Physical Review B}, 62\penalty0 (12):\penalty0 8410--8419, 2000.
\newblock ISSN 1098-0121.
\newblock \doi{10.1103/physrevb.62.8410}.

\bibitem[Foden et~al.(2000)Foden, Talyanskii, Milburn, Leadbeater, and Pepper]{Foden2000}
C.~L. Foden, V.~I. Talyanskii, G.~J. Milburn, M.~L. Leadbeater, and M.~Pepper.
\newblock {High-frequency acousto-electric single-photon source}.
\newblock \emph{Physical Review A}, 62\penalty0 (1):\penalty0 011803, 2000.
\newblock ISSN 1050-2947.
\newblock \doi{10.1103/physreva.62.011803}.

\bibitem[Rodriquez et~al.(2005)Rodriquez, Oi, Kataoka, Barnes, Ohshima, and Ekert]{Roberta2005}
Roberta Rodriquez, Daniel K.~L. Oi, Masaya Kataoka, Crispin H.~W. Barnes, Toshio Ohshima, and Artur~K. Ekert.
\newblock {Surface-acoustic-wave single-electron interferometry}.
\newblock \emph{Physical Review B}, 72\penalty0 (8):\penalty0 085329, 2005.
\newblock ISSN 1098-0121.
\newblock \doi{10.1103/physrevb.72.085329}.

\bibitem[Bäuerle et~al.(2018)Bäuerle, Glattli, Meunier, Portier, Roche, Roulleau, Takada, and Waintal]{Bäuerle2018}
Christopher Bäuerle, D~Christian Glattli, Tristan Meunier, Fabien Portier, Patrice Roche, Preden Roulleau, Shintaro Takada, and Xavier Waintal.
\newblock {Coherent control of single electrons: a review of current progress}.
\newblock \emph{Reports on Progress in Physics}, 81\penalty0 (5):\penalty0 056503, 2018.
\newblock ISSN 0034-4885.
\newblock \doi{10.1088/1361-6633/aaa98a}.

\bibitem[Edlbauer et~al.(2022)Edlbauer, Wang, Crozes, Perrier, Ouacel, Geffroy, Georgiou, Chatzikyriakou, Lacerda-Santos, Waintal, Glattli, Roulleau, Nath, Kataoka, Splettstoesser, Acciai, Figueira, Öztas, Trellakis, Grange, Yevtushenko, Birner, and Bäuerle]{Edlbauer2022}
Hermann Edlbauer, Junliang Wang, Thierry Crozes, Pierre Perrier, Seddik Ouacel, Clément Geffroy, Giorgos Georgiou, Eleni Chatzikyriakou, Antonio Lacerda-Santos, Xavier Waintal, D.~Christian Glattli, Preden Roulleau, Jayshankar Nath, Masaya Kataoka, Janine Splettstoesser, Matteo Acciai, Maria Cecilia da~Silva Figueira, Kemal Öztas, Alex Trellakis, Thomas Grange, Oleg~M. Yevtushenko, Stefan Birner, and Christopher Bäuerle.
\newblock {Semiconductor-based electron flying qubits: review on recent progress accelerated by numerical modelling}.
\newblock \emph{EPJ Quantum Technology}, 9\penalty0 (1):\penalty0 21, 2022.
\newblock ISSN 2662-4400.
\newblock \doi{10.1140/epjqt/s40507-022-00139-w}.

\bibitem[Kataoka et~al.(2006)Kataoka, Ford, Barnes, Anderson, Jones, Beere, Ritchie, and Pepper]{Kataoka2006}
M.~Kataoka, C.~J.~B. Ford, C.~H.~W. Barnes, D.~Anderson, G.~A.~C. Jones, H.~E. Beere, D.~A. Ritchie, and M.~Pepper.
\newblock The effect of pulse-modulated surface acoustic waves on acoustoelectric current quantization.
\newblock \emph{Journal of Applied Physics}, 100:\penalty0 063710, 2006.
\newblock \doi{10.1063/1.2349487}.

\bibitem[ota(2023)]{Ota2023}
S.~ota.
\newblock {On-Demand Single-Electron Source via Single-Cycle Acoustic Pulses}.
\newblock \emph{arXiv:2312.00274}, 2023.

\bibitem[Shilton et~al.(1996)Shilton, Mace, Talyanskii, Galperin, Simmons, Pepper, and Ritchie]{Shilton1996}
J.~M. Shilton, D.~R. Mace, V.~I. Talyanskii, Y.~Galperin, M.~Y. Simmons, M.~Pepper, and D.~A. Ritchie.
\newblock On the acoustoelectric current in a one-dimensional channel.
\newblock \emph{Journal of Physics: Condensed Matter}, 8:\penalty0 L337, 1996.
\newblock \doi{10.1088/0953-8984/8/24/001}.

\bibitem[Talyanskii et~al.(1997)Talyanskii, Shilton, Pepper, Smith, Ford, Linfield, Ritchie, and Jones]{Talyanskii1997}
V.~I. Talyanskii, J.~M. Shilton, M.~Pepper, C.~G. Smith, C.~J.~B. Ford, E.~H. Linfield, D.~A. Ritchie, and G.~A.~C. Jones.
\newblock Single-electron transport in a one-dimensional channel by high-frequency surface acoustic waves.
\newblock \emph{Physical Review B}, 56:\penalty0 15180, 1997.
\newblock \doi{10.1103/physrevb.56.15180}.

\bibitem[Cunningham et~al.(1999)Cunningham, Talyanskii, Shilton, Pepper, Simmons, and Ritchie]{Cunningham1999}
J.~Cunningham, V.~I. Talyanskii, J.~M. Shilton, M.~Pepper, M.~Y. Simmons, and D.~A. Ritchie.
\newblock Single-electron acoustic charge transport by two counterpropagating surface acoustic wave beams.
\newblock \emph{Physical Review B}, 60:\penalty0 4850, 1999.
\newblock \doi{10.1103/physrevb.60.4850}.

\bibitem[Cunningham et~al.(2000)Cunningham, Talyanskii, Shilton, Pepper, Kristensen, and Lindelof]{Cunningham2000}
J.~Cunningham, V.~I. Talyanskii, J.~M. Shilton, M.~Pepper, A.~Kristensen, and P.~E. Lindelof.
\newblock Single-electron acoustic charge transport on shallow-etched channels in a perpendicular magnetic field.
\newblock \emph{Physical Review B}, 62:\penalty0 1564, 2000.
\newblock \doi{10.1103/physrevb.62.1564}.

\bibitem[Ebbecke et~al.(2002)Ebbecke, Pierz, and Ahlers]{Ebbecke2002}
J.~Ebbecke, K.~Pierz, and F.~Ahlers.
\newblock Influence of the shape of a quasi-one-dimensional channel on the quantized acousto-electric current.
\newblock \emph{Physica E: Low-dimensional Systems and Nanostructures}, 12:\penalty0 466, 2002.
\newblock \doi{10.1016/s1386-9477(01)00336-8}.

\bibitem[P.~Utko and Lindelof(2003)]{Utko2003}
J.~B.~Hansen P.~Utko, K.~Gloos and P.~E. Lindelof.
\newblock {An Improved 2.5 GHz Electron Pump: Single-Electron Transport through Shallow-Etched Point Contac ts Driven by Surface Acoustic Waves}.
\newblock \emph{Acta Phys. Pol. A}, 103:\penalty0 533, 2003.

\bibitem[Ford(2017)]{Ford2017}
Christopher J.~B. Ford.
\newblock {Transporting and manipulating single electrons in surface‐acoustic‐wave minima}.
\newblock \emph{physica status solidi (b)}, 254\penalty0 (3):\penalty0 1600658, 2017.
\newblock ISSN 0370-1972.
\newblock \doi{10.1002/pssb.201600658}.

\bibitem[Morgan(2007)]{Morgan2007}
David Morgan.
\newblock \emph{{Surface acoustic wave filters: with applications to electronic communications and signal processing. 2nd ed.}}
\newblock Oxford: Academic Press, 2007.

\bibitem[Lima et~al.(2003)Lima, Alsina, Seidel, and Santos]{Lima2003}
M.~M.~de Lima, F.~Alsina, W.~Seidel, and P.~V. Santos.
\newblock {Focusing of surface-acoustic-wave fields on (100) GaAs surfaces}.
\newblock \emph{Journal of Applied Physics}, 94\penalty0 (12):\penalty0 7848--7855, 2003.
\newblock ISSN 0021-8979.
\newblock \doi{10.1063/1.1625419}.

\bibitem[Schülein et~al.(2015)Schülein, Zallo, Atkinson, Schmidt, Trotta, Rastelli, Wixforth, and Krenner]{Schülein2015}
Florian J.~R. Schülein, Eugenio Zallo, Paola Atkinson, Oliver~G. Schmidt, Rinaldo Trotta, Armando Rastelli, Achim Wixforth, and Hubert~J. Krenner.
\newblock {Fourier synthesis of radiofrequency nanomechanical pulses with different shapes}.
\newblock \emph{Nature Nanotechnology}, 10\penalty0 (6):\penalty0 512--516, 2015.
\newblock ISSN 1748-3387.
\newblock \doi{10.1038/nnano.2015.72}.

\bibitem[Ekström et~al.(2017)Ekström, Aref, Runeson, Björck, Boström, and Delsing]{Ekström2017}
Maria~K. Ekström, Thomas Aref, Johan Runeson, Johan Björck, Isac Boström, and Per Delsing.
\newblock {Surface acoustic wave unidirectional transducers for quantum applications}.
\newblock \emph{Applied Physics Letters}, 110\penalty0 (7):\penalty0 073105, 2017.
\newblock ISSN 0003-6951.
\newblock \doi{10.1063/1.4975803}.

\bibitem[Dumur et~al.(2019)Dumur, Satzinger, Peairs, Chou, Bienfait, Chang, Conner, Grebel, Povey, Zhong, and Cleland]{Dumur2019}
É. Dumur, K.~J. Satzinger, G.~A. Peairs, Ming-Han Chou, A.~Bienfait, H.-S. Chang, C.~R. Conner, J.~Grebel, R.~G. Povey, Y.~P. Zhong, and A.~N. Cleland.
\newblock {Unidirectional distributed acoustic reflection transducers for quantum applications}.
\newblock \emph{Applied Physics Letters}, 114\penalty0 (22):\penalty0 223501, 2019.
\newblock ISSN 0003-6951.
\newblock \doi{10.1063/1.5099095}.

\end{thebibliography}






\end{document}